\documentclass[manuscript]{emulateapj}

\usepackage{graphicx}
\usepackage{graphpap}
\usepackage{epsf}
\usepackage{amssymb}
\usepackage{color}

\def\rpp{Reports of Progress in Physics}%
          \newcommand{\kms}{{km\,s$^{-1}$}}

\slugcomment{Not to appear in Nonlearned J., 45.}

\shorttitle{H$\alpha$ line emission of SNR 0509-67.5}
\shortauthors{Helder, Kosenko and Vink}

\begin{document}


\title{Cosmic-ray acceleration efficiency vs temperature equilibration: \\the case of SNR 0509-67.5  }


\author{E. A. Helder\altaffilmark{1}, D. Kosenko\altaffilmark{1} and J. Vink\altaffilmark{1}}
\affil{$^{1}$Astronomical Institute Utrecht, Utrecht University, P.O. Box 80000, NL-3508 TA Utrecht, The Netherlands}
\email{e.a.helder@astro-uu.nl}

\begin{abstract}
 {We study the 0509-67.5 supernova remnant in the Large Magellanic Cloud with the VLT/FORS2 spectrograph. We detect a broad component in the H$\alpha$ emission with a FWHM of $2680\pm 70$ \kms\ and $3900 \pm 800$ \kms\ for the southwest (SW) and northeast (NE) shocks respectively. For the SW, the proton temperature appears to be too low for the shock velocity, which we attribute to a cosmic-ray pressure behind the shock front of at least 20\% of the total pressure. For the NE, the post-shock proton temperature and the shock velocity are compatible, only if the plasma behind the shock front has a degree of thermal equilibrium of over 20\%, which is at odds with current models for temperature equilibration behind fast shocks, which do not accelerate cosmic rays. If we assume the electron temperature to be less than 10\% of the proton temperature, we find a post-shock cosmic-ray pressure of at least 7\%.}
\end{abstract}

\section{Introduction}
Supernova remnants (SNRs) are generally thought to be the dominant sources of Galactic cosmic rays. For this to be true, SNRs need to transfer about 10\% of their initial kinetic energy into cosmic rays. An open question is whether SNR shocks can reach these acceleration efficiencies averaged over their life time; recent TeV and GeV $\gamma$-ray observations give promising but ambiguous results \citep{Ellison2010,Abdo2010}.

Although the process of efficiently accelerating particles is well understood \citep{Malkov}, observational verifications for efficient acceleration are scarce \citep[e.g. ][]{Warren2005, Lee2007, Vink2006, Helder2009}. These observations are essential for characterizing the efficiency of the acceleration. As the particles move ahead (i.e. upstream) of the shock, they form a so-called shock precursor, which compresses and pre-heats the upstream medium. This effectively lowers the Mach number of the main shock, therewith lowering the temperature behind the shock front \citep{Ellison2004, Drury2009}. This effect would ideally be characterized by the post-shock proton temperature ($T_{\rm s,\,p}$), as $T_{\rm s,\,p}$ is close to the plasma temperature, whereas the post-shock electron temperature $T_{\rm s,\,e}$ might be lower \citep[e.g. ][]{Ghavamian2007}. The latter implies that the electrons might constitute only a minor part of the thermal pressure behind the shock front. 

For some SNRs, $T_{\rm s,\,p}$ can be determined from hydrogen line emission at the shock fronts; these are so-called Balmer-dominated shocks. The hydrogen lines of a Balmer-dominated shock consist of two superimposed components; the narrow component is emitted by neutral hydrogen after entering the shock front and the broad component by hot protons after undergoing charge exchange with incoming neutral hydrogen atoms. The width of the broad component reflects the proton temperature behind the shock front \citep{Chevalier1980, Heng2009}. 
\begin{figure*}[!t]
\begin{center}
\begin{tabular}{c c c}
\includegraphics[angle = 0, width=0.33\textwidth]{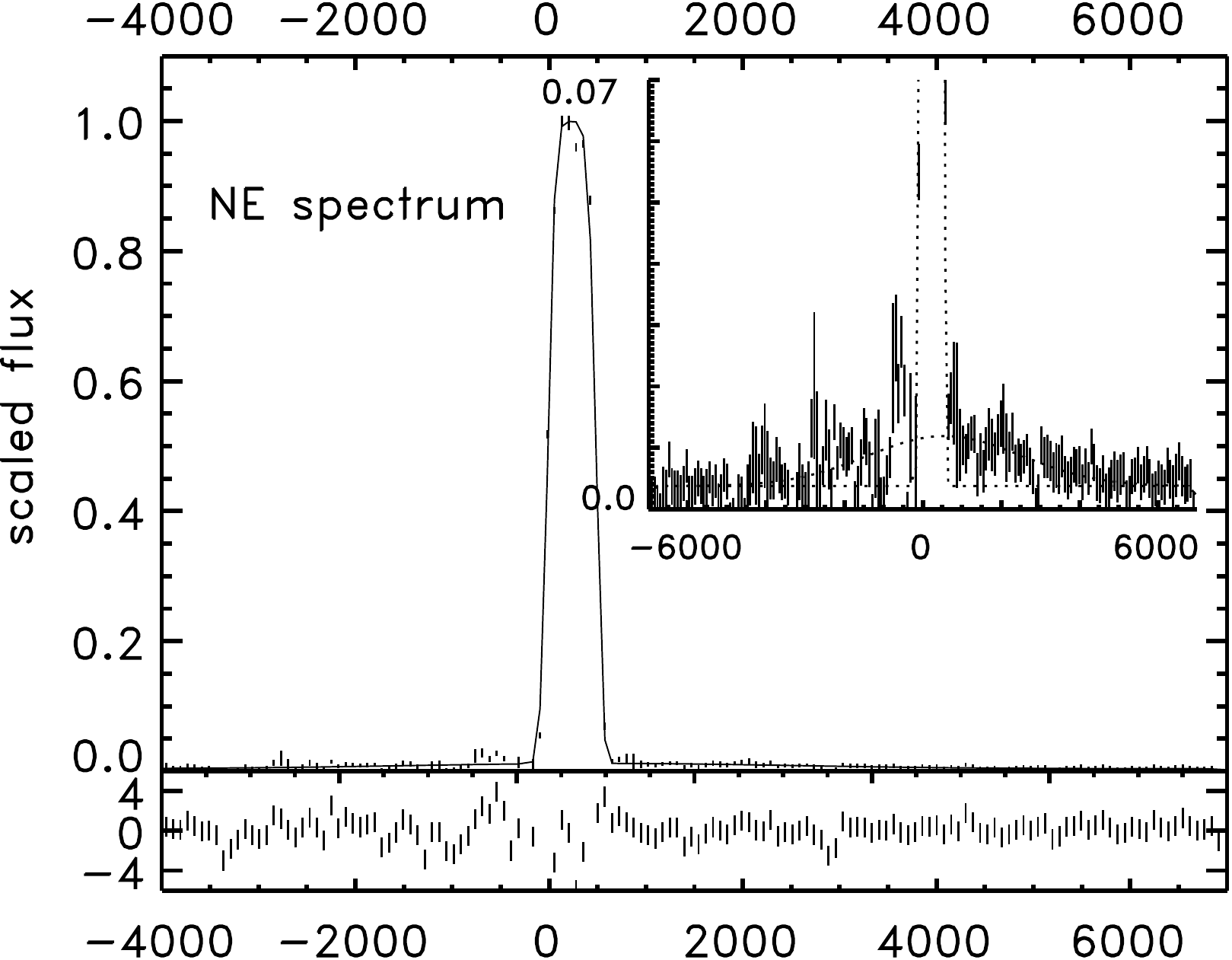} &
\includegraphics[angle = 0, width=0.24\textwidth]{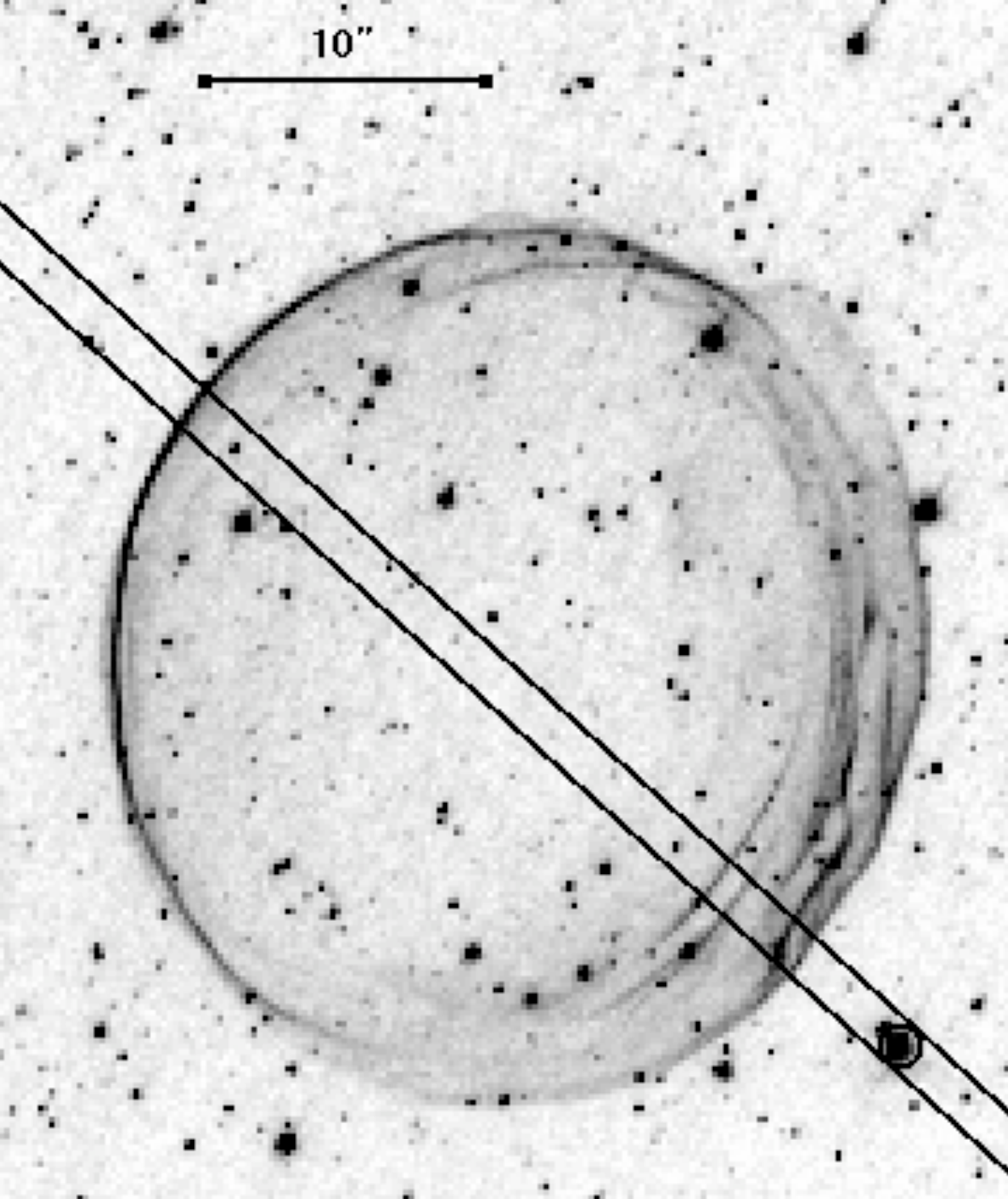} & 
\includegraphics[angle = 0, width=0.33\textwidth]{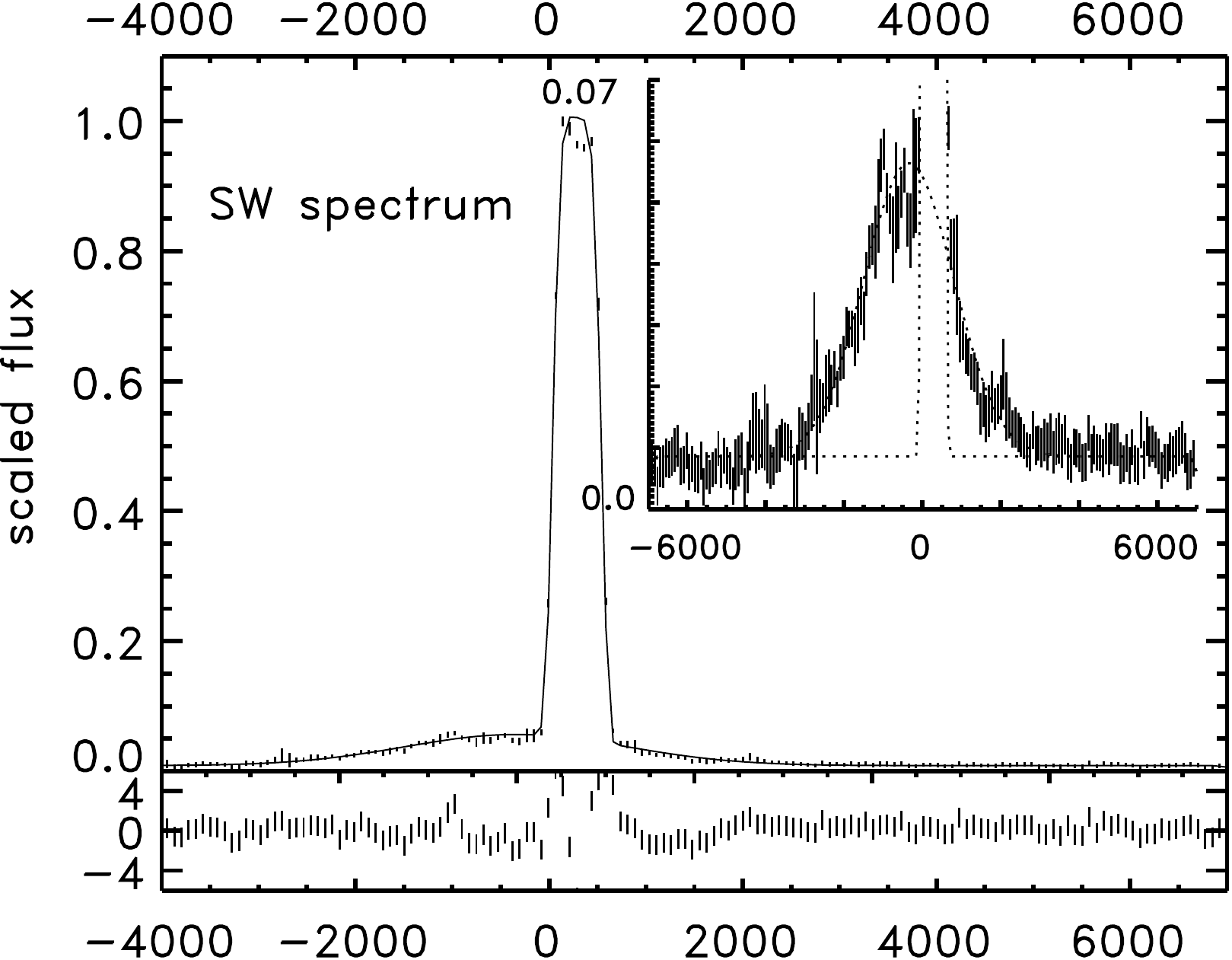} \\
\end{tabular}
\caption{{\em Left:} Spectrum and best fit of the NE shock of SNR 0509-67.5. {\em Middle:} Archival HST/ACS observation of SNR 0509-67.5, as obtained at October 28, 2006, with an exposure time of 4620 seconds. The position of the longslit of the VLT/FORS2 instrument is indicated. The star on which the slit is centered is indicated with a circle. North is up, east is to the left. {\em Right: } Spectrum and best fit of the SW shock. The lower panels show the residuals divided by the errors. The insets show a zoom-in of the broad component, with the best fit narrow and broad component overplotted.}
	\vskip 7mm
              \label{HST}%
\end{center}
\end{figure*}

In this {\it Letter}, we report on our study of SNR 0509-67.5, as this is a likely source of efficient cosmic-ray acceleration; recent studies show that the remnant resulted from a highly energetic Type Ia explosion \citep{Hughes1995, Badenes2008, Rest2008}. Its X-ray spectrum provides some evidence for non-thermal emission \citep{Warren2004}. In addition, hydrodynamical models show that the presence of highly energetic particles is likely not negligible \citep{Kosenko2008}. 

SNR 0509-67.5 was discovered to have Balmer-dominated shocks by \cite{Tuohy}, although the flux in the broad component was too low to be detected. A subsequent attempt by \cite{Smith1991} did not reveal any broad component either. The first detection of a broad component in the hydrogen line emission of this remnant was by \cite{Ghavamian2007b}, who measured the width of the broad component of the Ly$\beta$ line to be 3700 $\pm 400$~\kms, corresponding to a shock velocity ($v_{\rm s}$) of 5200-6300 \kms. However, as the spectrum was taken from the entire remnant, it remains uncertain whether the obtained linewidth is broadened by the bulk motion of the plasma as well. 

We report the detection of a broad component in the H$\alpha$ line emission of SNR 0509-67.5 at two locations of the shock front. We combine this with a shock velocity based on X-ray observations \citep{Kosenko2008}, to determine the fraction of the post-shock pressure contributed by cosmic rays.

\section{Optical data and results}\label{data}
We observed SNR 0509-67.5 for 10932 seconds (4 blocks of $2733$ seconds) with FORS2, the low dispersion spectrograph of ESO's VLT \citep{FORS}. The observations were made on October 15 and 20 and November 10 and 11, 2009. By centering the slit at a bright star at $\alpha = 05:09:28.793$, $\delta = -67:31:30.83$ (J2000), with a position angle of 42$^\circ$ (Figure \ref{HST}), we obtained spectra of both the SW and NE rims with a single pointing. The width of the slit was 1.6$\arcsec$ which, in combination with the 600RI grism and the $2\times 2$ binned readout, corresponds to a resolution of $\sim$ 485 \kms~at H$\alpha$ (6563 \AA). This resolution prevents us from resolving the narrow component of the H$\alpha$ line, but it increases the signal-to-noise sufficiently to detect and resolve the broad component. The data were corrected for bias, flatfielded, and the skylines and cosmic rays were removed. The wavelength calibration was done by fitting a fourth order polynomial to the spectral lines of He, HgCd, Ar and Ne lamps, obtained during daytime. In addition, we checked our calibration against the position of 3 skylines \citep[at 5577.3, 6864.0 and 7571.7 \AA\ respectively, ][]{Osterbrock1996}. 
The absolute wavelength calibration has a systematic uncertainty of 0.5 pixel ( 37 \kms) and the errors on the datapoints were determined by tracing the error propagation through the reduction steps, starting from the raw data. For each observation block, obvious outliers, probably caused by erroneous skyline subtraction, were removed. The resulting spectra were fitted with two superimposed Gaussians and an offset, convolved with the slit width. 
The best fitting parameters for both spectra are listed in Table \ref{fit}. In order to determine the significance of the broad component in the NE spectrum, we compared the obtained $\chi^2$ value (464) with a fit in which we only fitted a single Gauss to the spectrum (525). The difference indicates a significance of 7.8$\sigma$. The $\chi^2_{\rm red}$ values are 2.03 and 2.56 for the SW and NE fits respectively. These high values are mainly caused by substructure of the narrow line: a fit to the broad component with a single Gauss, excluding the central region (between -100 and 800 \kms), convolved with the resolution, gives $\chi^2_{\rm red}$ of 0.51 for the SW and 0.76 for the NE. This substructure might either be caused by spatial surface brightness fluctuations of the remnant within the slit, or by errors in the skyline subtraction. We determine the $1\sigma$ errors on the parameters for the broad components by using the errors we determined in our data reduction. For the narrow component we first increased the errors such that  $\chi^2_{\rm red}=1$, and then determined the 1 $\sigma$ errors, using $\Delta \chi^2=1$.

\begin{table*}
\begin{center}
\begin{tabular}{lccc}
Parameter&  & SW & NE    \\ 
 \hline 
center narrow& [\kms]& $287.0 \pm 1.4$ & $286.0\pm 1.5$\\
center broad &[\kms]& $-342\pm 28$ & $ 459\pm 220$ \\ 
FWHM broad &[\kms] & $2680\pm 70 $ & $ 3900 \pm 800$  \\ 
total flux  & [10$^{-16}$ erg s$^{-1}$ cm$^{-2}$ arcsec$^{-2}$] & 8.6 & 5.3 \\
$I_{\rm b}/I_{\rm n}$ & & $0.29\pm 0.01$ & $0.08\pm 0.02$\\  
\hline
\end{tabular}
\end{center}
\label{fit}
\begin{list}{}{}
\item[$^\mathrm{a}$] Approximate flux calibration based on an observation of a photometric standard star \citep[LTT 2415,][]{Hamuy} taken on November 11, 2010
\end{list}
\end{table*}
\vskip 4mm

\section{Comparing with shock velocities}\label{RGS}
To determine the shock velocity, we use the X-ray line width of $\sigma_{\rm RGS} = 4900 \pm 420$ \kms, as observed by the Reflection Grating Spectrometer (RGS) onboard XMM-Newton \citep{Kosenko2008}. This line width is caused by both the thermal and bulk broadening of the plasma. We follow the method of \cite{Kosenko2010} to disentangle the bulk broadening from the thermal broadening to obtain the forward shock velocity, as described below.

As shown in \cite{Kosenko2010}, $\sigma_{\rm RGS}$ and $v_{\rm s}$ are related as follows:
\begin{equation}
\sigma_{\rm RGS} = (v_{\rm s}/4)\sqrt{3r_{\rm sh}^2+ 9r_{\rm bulk}^2},
\label{vRGS}
\end{equation}
where $r_{\rm sh} = v_{\rm rs}/v_{\rm s}$: the ratio ot the reverse shock velocity to the forward shock velocity and $r_{\rm bulk}=v_{\rm bulk,\ ej}/v_{\rm bulk,\ CSM}$: the gradient in the plasma bulk velocity from reverse to forward shock. 
We obtain $r_{\rm sh}  \simeq 1$ (whole remnant) and $r_{\rm sh} \simeq 0.5$ (NE), from analytical models \citep{Truelove} with 
$M_{\rm ej} = 1.4\ \rm{M_\odot}$, $E = 1.4\times10^{51}$ erg and an age of 400 years \citep{Badenes2008, Rest2008}. 
Additionally, we constrain the models with a forward shock radius of $15.9 \pm 0.8 \arcsec$ and $16.3 \pm 0.3 \arcsec$ for the entire remnant and the NE respectively and a reverse shock radius of $11.4 \pm 2.1\arcsec$ and $13.5 \pm 0.2\arcsec$ for the entire remnant and the NE respectively, based on a deprojection of Chandra images (48.9 ks, obs ID:776), following the procedure of \cite{Kosenko2010}. 

We obtain $r_{\rm bulk} = 0.95$ with numerical simulations \citep{Sorokina2004, Kosenko2006}, using the above parameters. 

We use equation (\ref{vRGS}) to estimate $v_{\rm s} = 6000\pm300$ \kms\ (whole remnant) and $v_{\rm s} = 6600\pm 400$ \kms\ (NE). 
In the remainder of this paper, we use conservative $v_{\rm s}$ estimates of 5000 \kms\ for the SW and 6000 \kms\ for the NE.

To check these shock velocities, we estimated the expansion with Chandra. Chandra observed SNR 0509-67.5 for 3 times, with the first observation in May 2000 (earlier used in this section to determine forward and reverse shock radii) and the latter two in May 2007 (29.5 and 32.7 ks, obs IDs 7635 and 8554). Following the method of \cite{Vinkkepler}, we find a shock velocity of 6700 $\pm 400$ \kms\  averaged over the azimuth of the remnant. Note that a full proper motion study of SNR 0509-67.5, using both Chandra and Hubble data, is underway \citep{Hughes2010}. The dense material in the SW (Fig. \ref{HST}) suggests that the forward shock might have recently slowed down. However, we do not find any support for this scenario from the Chandra expansion study, nor from our XMM-Newton/RGS study. Moreover, as the RGS study is skewed towards the bright SW region, it is unlikely that the SW velocity is substantially lower than 6000 \kms.  \\

\section{Interpretation}

The centroid of the broad component in the SW of $-342\pm28$ \kms\ indicates the bulk line-of -sight velocity of the shocked protons. Additionally, the Large Magellanic Cloud (LMC) is moving with a 278 \kms\ away from us \citep{Richter1987}. This means that we are observing a part of the shell which is moving toward us with 620 \kms\ with respect to the LMC. 
Figure \ref{HST} shows  a bright, inner shell close to the outer shock. Taking into account the seeing of 0.8-1.0\arcsec, our spectrum is probably contaminated by emission from this shell. This part of the shell is likely to dominate the measurement by \cite{Ghavamian2007b}, as they also measured a positive offset for the centroid of the broad component.

The flux in the broad component with respect to the flux in the narrow component ($I_{\rm b}/I_{\rm n}$) declines as a function of shock velocity \citep{Heng2007,Adelsberg}. Our low values for $I_{\rm b}/I_{\rm n}$ are therefore consistent with a high shock velocity with likely the highest shock velocity in the NE, which has the lowest $I_{\rm b}/I_{\rm n}$. 
However, as the standard models for interpreting Balmer dominated shocks do not include the effects of cosmic-ray acceleration, it is not appropriate to use the $I_{\rm b}/I_{\rm n}$ values to determine the shock velocities of SNR 0509-67.5.

\begin{figure}[!b]
\begin{center}
\includegraphics[angle = 0, width=0.45\textwidth]{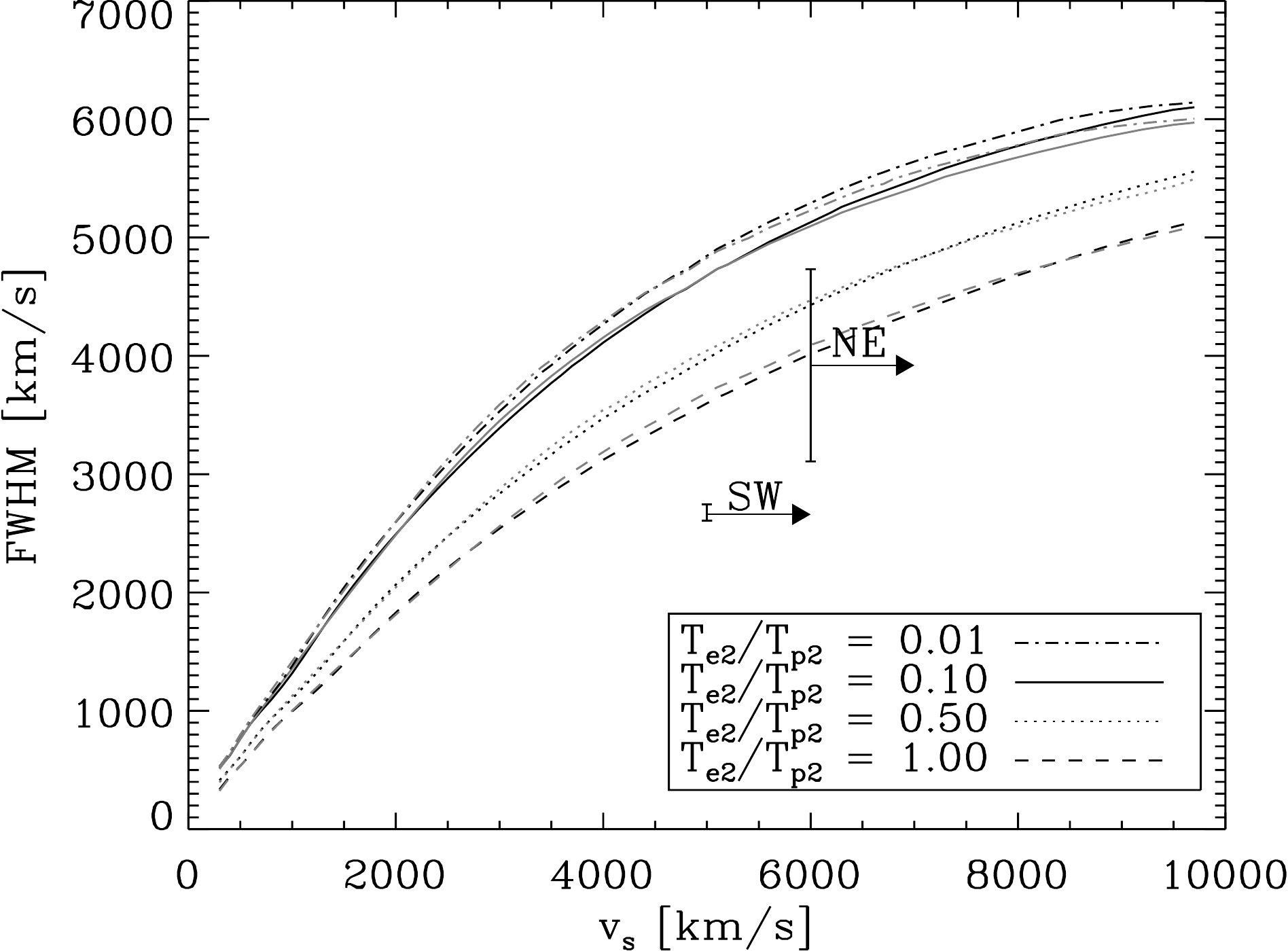}  \\
\caption{FWHM as function of $v_{\rm s}$ for non-accelerating shocks, for different values for $T_{\rm e}/T_{\rm p}$, (courtesy to K. Heng and M. van Adelsberg). The grey (black) lines indicate this relation for plasmas, that are optically thin (thick) to Ly$\beta$ scattering \citep{Adelsberg}. Overplotted is $v_{\rm s}$ and the measured FWHM for both the SW and NE. } 
             \label{kosenkadel}%
\end{center}
\end{figure}

\begin{figure}[!b]
\begin{center}
\includegraphics[angle = 0, width=0.55\textwidth]{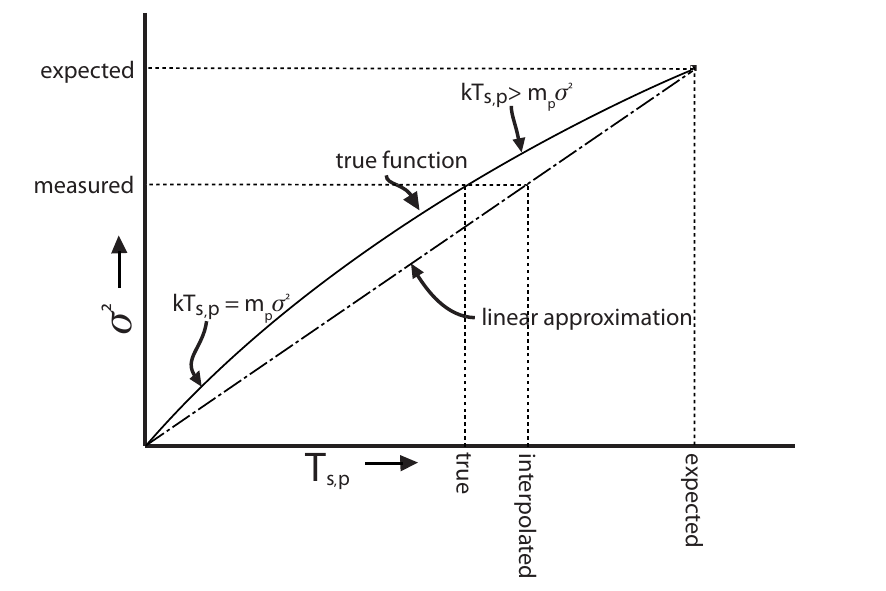}  \\
\caption{Cartoon of $\sigma^2$ as function of $T_{\rm s,\,p}$. `Expected' are the values we expect for a non-accelerating shock (Fig. \ref{kosenkadel}). `True' shows the true $T_{\rm s,\,p}$ and `measured' is the corresponding $\sigma^2$ we would measure. `Interpolated' shows our estimated $T_{\rm s,\,p}$, based on the measured $\sigma^2$. Note that this method leads to an overestimation of $T_{\rm s,\,p}$. } 
             \label{protont}%
\end{center}
\end{figure}

\subsection{Interpreting the FHMWs}\label{interpretation}
To determine the cosmic-ray pressure from our measurements, we need to determine $T_{\rm s,\,p}$ from the FWHM. For low temperatures and shock velocities, $kT_{\rm s,\,p} = m_{\rm p} \sigma^2$ \citep{Rybicki}, with $\sigma = \rm{FWHM}/\sqrt{8\ln{2}}$ in ${\rm cm~s}^{-1}$. However, as cross sections for charge exchange decline for high velocities \citep[e.g. Fig. 1 in][]{Heng2007}, $kT_{\rm s,\, p} > m_{\rm p}\sigma^2$ for higher temperatures and shock velocities. Recent studies focused on determining the FWHM as function of $v_{\rm s}$ for non-accelerating shocks \citep[e.g. ][resulting in Fig. \ref{kosenkadel}]{Heng2007, Adelsberg}. 
Here, we are interested in FWHM as function of $T_{\rm s,\,p}$ for a given $v_{\rm s}$ instead. We approximate this concave function linearly between 0,0 and the FWHM and $T_{\rm s,\,p}$ expected for a non-accelerating shock in thermal equilibrium (Fig. \ref{kosenkadel} and \ref{protont}). In this way, we overestimate $T_{\rm s,\,p}$ and the corresponding thermal pressure, leading to a conservative measure of the cosmic-ray pressure behind the shock front. 

We note that the y-axis of Fig. 5 of van \cite{Adelsberg} is labeled incorrectly (M. van Adelsberg, 2010, private communication). Instead of showing the `broad H$\alpha$ FWHM', this figure plots the `broad neutral velocity distribution FWHM', which is independent of the emission line considered. To model a specific emission line such as H$\alpha$, one has to convolve the broad neutral velocity distribution with the relevant atomic cross sections.  The FWHM-$v_{\rm s}$ relations we use in this study, obtained in electronic form from M. van Adelsberg and K. Heng, are plotted in Fig. \ref{kosenkadel} and are based upon the same relations used to generate Figure 13 and Table 1 of \cite{Adelsberg}.

Following \cite{Vinkreview2008} and \cite{Helder2009}, we interpret $T_{\rm s,\,p}$ and $v_s$ in terms of cosmic-ray pressure behind the shock front and cosmic-ray energy flux leaving the system (respectively $w_{\rm CR} = P_{\rm CR}/P_{\rm tot}$ in which $P_{\rm tot}$ is the total pressure behind the shock front (x-axis in Figure \ref{energy}) and $\epsilon_{\rm esc} = F_{\rm esc}/F_{\rm tot}$ in which $F_{\rm tot}$ is the energy flux entering the shock; $\frac{1}{2}\rho_{\rm ISM}v_{\rm s}^3$ (y-axis of Figure \ref{energy})). To conservatively estimate the post-shock cosmic-ray pressure, we assume the electrons and ions to be in thermal equilibrium. We add $w_{\rm CR}$ and $\epsilon_{\rm esc}$ to the equations of conservation of mass, momentum and energy over the shock front, which leads to:
\begin{eqnarray}
kT_{\rm s,\,p}=(1-w_{\rm CR})\frac{1}{\chi}(1 - \frac{1}{\chi})\mu m_{\rm p}v_{\rm s}^2.
\label{heatCR}
\end{eqnarray}
Formally, $\mu$ is the number averaged mean particle weight, ($\sim$ 0.54 for a fully equilibrated and fully ionized plasma with LMC abundances), when considering $kT_{\rm s,\,p}$, we can treat $\mu$ as well as a measure for the temperature equilibration behind the shock front, where $\mu = 1$ indicates no temperature equilibration, and $\mu = 0.54$ indicates a fully equilibrated plasma. In addition, $\chi$ is the total shock compression ratio. 
For a non-accelerating, adiabatic shock, $\chi = 4$ and hence $kT_{\rm s,\,p}=\frac{3}{16}\mu m_{\rm p}v_{\rm s}^2$. We define $\beta \equiv kT_{\rm s,\,p}/\frac{3}{16}\mu m_{\rm p}v_{\rm s}^2$, to characterize the influence of cosmic-ray acceleration on $T_{\rm s,\,p}$. Figure \ref{energy} shows $\beta$ in the ($w_{\rm CR}$, $\epsilon_{\rm esc}$)-frame.  The `max'-line indicates the ratio of the cosmic-ray pressure and escaping cosmic-ray energy for the most efficiently accelerating shock according to theory \citep{Malkov1999,Drury2009}. Hence, the hashed region is excluded. 


\subsection{The southwest spectrum}

For $v_{\rm s} =$ 5000 \kms, we expect to measure at least a FWHM for the broad component of 3600 \kms , corresponding to a $T_{\rm s,\,p}$ of 28.7 keV (Fig \ref{kosenkadel}). This differs significantly from the measured $2680 \pm 70$~\kms. We derive a $T_{\rm s,\,p}$ of $15.9\pm 0.9$ keV, leading to a $\beta < 0.58$, which gives a cosmic-ray pressure behind the shock front of at least 20\% (Fig. \ref{energy}).

\subsection{The northeast spectrum}
\begin{figure}[!t]
\begin{center}
\includegraphics[angle = 0, width=0.46\textwidth]{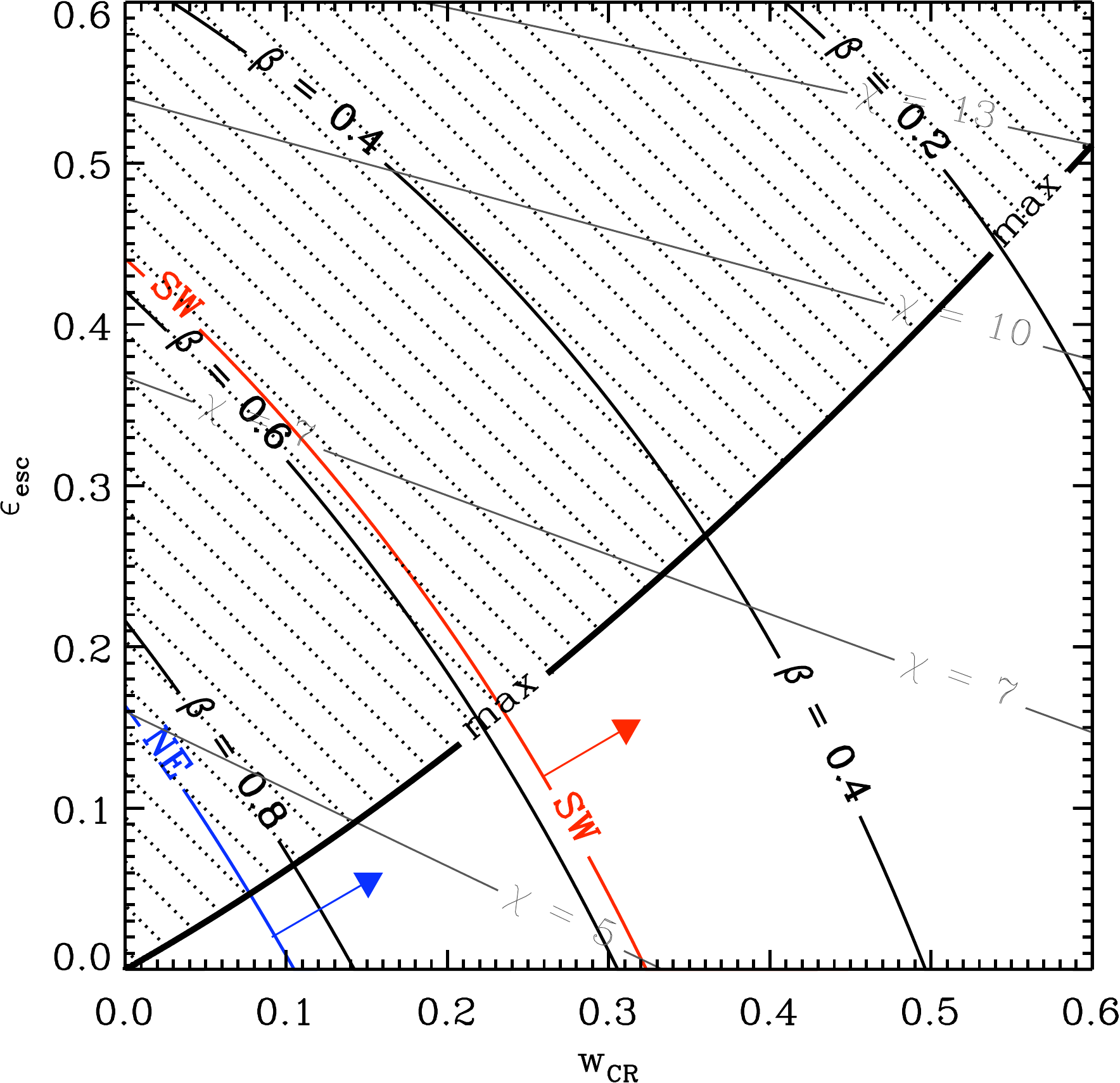}  \\
\caption{Values for $\beta$ in the ($w_{\rm CR}, \epsilon_{\rm esc}$)-frame. The $\chi$-lines indicate the compression ratio of the plasma behind the shock front. The hashed region requires a shock, more efficient than can be explained with current theoretical models. The red line indicates the lower limit for $\beta$ for the SW. For the NE, $\beta$ is determined assuming $T_{\rm s,\,e}/T_{\rm s,\,p} <0.1$.
} 
              \label{energy}%
\end{center}
\end{figure}

For the NE, the measured FWHM indicates a $v_{\rm s}=$ 6000 \kms, for 0.2$<T_{\rm s,\,e}/T_{\rm s,\, p}<1.0$  (Fig. \ref{kosenkadel}). This leads to two possibilities: \\
1) The shock does not efficiently accelerate cosmic rays and has a $T_{\rm s,\,e}/T_{\rm s,\, p}>0.2$, breaking with the earlier reported trend of $T_{\rm s,\,e}/T_{\rm s,\,p}\propto 1/v_{\rm s}^2$ for $v_{\rm s}>400$ \kms\ \citep{Ghavamian2007b}, as this would give $T_{\rm e}/T_{\rm p} < 0.01$ for $v_{\rm s} = 6000$ \kms.\\
 2) The shock is far out of thermal equilibrium, as we might expect for a fast shock, and is accelerating particles. If we assume $T_{\rm s,\,e}/T_{\rm s,\,p} < 0.1$, we expect a FWHM of $>$5100 \kms, whereas we measure $3900 \pm 800$ \kms. Using the approach from section \ref{interpretation} and Fig. \ref{energy}, we obtain $\beta < 0.85$ and hence a cosmic-ray pressure behind the shock front of $>7 \%$ of the total pressure. Note that this is a very conservative lower limit: for both $T_{s,\,p}$ and $v_{\rm s}$ we used conservative approximations. Also, at this high shock velocity, the squared line width of the broad component does not increase linearly with $T_{\rm s,\,p}$, but flattens off. This makes the difference in line width lower for a given $\Delta T_{\rm s,\, p}$, which makes that the line width is a less sensitive temperature indicator at high shock velocities. 
 Future progress can be made with better shock velocity estimates of the NE region itself, a higher signal-over-noise spectrum of the NE region and with models for non-radiative Balmer dominated shocks that include the effects of cosmic-ray acceleration.

A remaining question is whether the magnetic field pressure makes a contribution to the post-shock pressure. As SNR 0509-67.5 is at a distance of 50 kpc, we can not resolve filaments of X-ray synchrotron emission, which are often used for determining post-shock magnetic field strengths \citep{Vink2003}. However, \cite{Voelk2005} showed that a typical value for the magnetic field pressure is around 3.5\% of the total post-shock pressure. Moreover, according to Bell's theory \citep{Bell2004} the magnetic field energy density scales as
$B^2/8\pi \sim \frac{1}{2} v_{\rm s}U_{\rm c} /c $, with $U_{\rm c}$ the cosmic ray energy density. This means that the magnetic field energy density is expected to be about 1\% of the cosmic-ray energy density.

\section{Conclusions}
We investigated the cosmic-ray acceleration efficiency of the 0509-67.5 SNR in the LMC, by comparing  $T_{\rm s,\,p}$, determined from the H$\alpha$-line widths of the SW and NE shocks with shock velocities of respectively 5000 \kms\ and 6000 \kms\ for the SW and NE, based on X-ray observations. Our study gives the following results:

\begin{itemize}
\renewcommand{\labelitemi}{-}
\item We measured widths of the broad components of the H$\alpha$ lines in SNR 0509-67.5 to be $2680 \pm 70$ \kms\ for the SW and $3900 \pm 800$ \kms\ for the NE.

\item For the SW, we can only explain the FWHM if we allow for a contribution of $>20\%$ of the post-shock pressure by cosmic rays.
\item For the NE shock,  we have two options: either the shock has a  $T_{\rm s,\,e}/T_{\rm s,\,p}>0.2$, breaking with the earlier reported trend of $T_{\rm s,\,e}/T_{\rm s,\,p}\propto 1/v_{\rm s}^2$ for $v_{\rm s} > 400$ \kms\ \citep{Ghavamian2007b}, or if  we assume $T_{\rm s,\,e}/T_{\rm s,\,p}<0.1$, the cosmic-ray pressure behind the shock front is at least 7\% of the total pressure. 

\end{itemize}

This research, together with our previous study \citep{Helder2009}, shows that more than 10\% of the pressure in young SNRs can be contributed by cosmic rays. This is more than the requirement that 10\% of the available energy needs to be in cosmic rays. On the other hand, the cosmic-ray acceleration efficiency may decline for older SNRs, as indicated by a recent study of the Cygnus Loop \citep{Salvesen2009}. So a higher efficiency at a young age may be needed to have an average efficiency of 10\% over the whole lifetime of a SNR.

\section{Acknowledgments}
The authors thank Matthew van Adelsberg and Kevin Heng for kindly providing the data used for Figure 2. We also thank Frank Verbunt for critical reading of the manuscript and Cees Bassa and Peter Jonker for useful discussions on the planning and reduction of optical spectra. This research is based on observations collected with ESO Telescopes at the Paranal Observatory under programme ID 384.D-0518(A) and on observations obtained with the XMM-Newton and Chandra satellites. E.A.H. and J.V. are supported by the Vidi grant of J.V. from the Netherlands Organization for Scientific Research (NWO). D.K. is supported by an ``open competition'' grant from NWO. This research has made use of the NASA/IPAC Extragalactic Database (NED) which is operated by the Jet Propulsion Laboratory, California Institute of Technology, under contract with the National Aeronautics and Space Administration.

\end{document}